\documentclass{webofc}
\usepackage[varg]{txfonts} 
\usepackage{graphicx,epsf,amsmath,amsfonts,amssymb,amsbsy}
\usepackage{epsfig}
\usepackage{epstopdf}
\usepackage{tikz}
\usepackage{verbatim}
\usepackage{textalpha}

\newcommand{\ds}{\displaystyle}
\newcommand{\vev}[1]{\langle#1\rangle}
\newcommand{\mat}{\left ( \begin{array}}
\newcommand{\emat}{\end{array} \right )}
\newcommand{\vect}{\left ( \begin{array}{c}}
\newcommand{\evect}{\end{array} \right )}

\begin{document}
\title{





Affinity of NJL$_2$ and NJL$_{4}$ model results on duality and pion condensation in chiral asymmetric dense quark matter

}
%
%

\author{\firstname{T. G.} \lastname{\it Khunjua}\inst{1}\fnsep\thanks{\email{gtamaz@gmail.com}} \and
        \firstname{K.G.} \lastname{\it Klimenko}\inst{2}\fnsep\thanks{\email{Konstantin.Klimenko@ihep.ru}} \and
        \firstname{R. N.} \lastname{\it Zhokhov--\it Larionov}\inst{2,3}\fnsep\thanks{\email{zhokhovr@gmail.com}}
}

\institute{ Faculty of Physics, Moscow State University,
119991, Moscow, Russia
\and
           Logunov Institute for High Energy Physics, NRC "Kurchatov Institute
\and
          Pushkov Institute of Terrestrial Magnetism, Ionosphere and
Radiowave Propagation (IZMIRAN), 108840 Troitsk, Moscow, Russia
          }

\abstract{
  In this paper we investigate the phase structure of a (1+1) and (3+1)-dimensional quark model with four-quark interaction and in the presence of baryon ($\mu_B$), isospin ($\mu_I$) and chiral isospin ($\mu_{I5}$) chemical
potentials. 
It is shown that the chemical potential $\mu_{I5}$ promotes the appearance of the charged PC phase with nonzero baryon density.
Results of both models are qualitatively the same, this fact enhances  one's confidence in 
the obtained predictions.
It is established that in the large-$N_c$ limit ($N_c$ is
the number of colored quarks) there exists a duality correspondence between the chiral symmetry breaking phase and the charged pion condensation one.
}
\maketitle

\section{Introduction}
Recently, much attention has been paid to the investigation of
the QCD phase diagram in the presence of baryonic as well as isotopic
(isospin) chemical potentials. The reason is that dense baryonic
matter which can appear in heavy-ion collision experiments has an
evident isospin asymmetry. Moreover, the dense hadronic/quark matter
inside compact stars is also expected to be isotopically asymmetric.

 However, theoretical investigations of QCD encounter considerable
difficulties in the low-energy as well as low-temperature and density region, where perturbative methods do not work.
The only possible first principle calculation in QCD at low energies is lattice QCD. Unfortunately, its
main method (Monte Carlo simulations) cannot be applied at finite baryon chemical potential due to
the sign problem. In order to study the phase diagram of QCD at nonzero chemical
potential one usually use effective field theories.
 One of the most widely used effective theory is Nambu–JonaLasinio
(NJL) model \cite{njl} (see for review \cite{Klevansky:1992qe, Hatsuda:1994pi, Buballa:2003qv}). The model
is tractable and can be used as low energy effective theory for QCD. In this way, QCD phase diagrams
including chiral symmetry restoration \cite{asakawa,ebert,sadooghi
}, color superconductivity \cite{alford,klim,incera}, and charged pion condensation (PC) phenomena \cite{
eklim,ak,mu,andersen} were investigated under heavy-ion experimental and/or compact star conditions, i.e. in the presence of temperature, chemical potentials and possible external (chromo)magnetic fields. 
(3+1)-dimensional NJL models are non-renormalizale and depend on the cutoff
parameter which is typically chosen to be of the order of 1 GeV, so that
the results of their usage are valid only at {\it comparatively low
energies, temperatures and densities (chemical potentials)}.
But there exists also a class of renormalizable theories, the
(1+1)-dimensional chiral Gross--Neveu (GN) or NJL type models 
,
 that can be used as a laboratory for the qualitative simulation of
specific properties of QCD at {\it arbitrary energies}.
Renormalizability, asymptotic freedom, as well as the spontaneous
breaking of chiral symmetry (in vacuum) are the most fundamental
inherent features both for QCD and all NJL$_2$ type models. In addition,
the $\mu_B-T$ phase diagram is qualitatively the same for the QCD and NJL$_2$ models
\cite{wolff,kgk1,
chodos}. 

Among all the above mentioned phenomena, which can be observed in dense baryonic matter, the existence of the charged PC phase is predicted without sufficient certainty. If the electric charge neutrality constraint is imposed,  the charged pion condensation phenomenon  depends strongly on the bare (current) quark mass
values. In particular, it turns out that the charged PC phase with
{\it nonzero baryonic density} is forbidden in the framework of NJL models for the physically acceptable
values of the bare quark masses (see ref. \cite{andersen}). Due to these circumstances, the question arises whether there exist factors promoting the appearance of charged PC
phenomenon in  dense baryonic matter. A positive answer to this
question was obtained in the papers \cite{ekkz,gkkz}, where it was
shown that a charged PC phase might be realized in dense baryonic system with finite
size or in the case of a spatially inhomogeneous pion condensate.

In the present paper we will show that a chiral imbalance of dense and isotopically asymmetric baryon matter is another interesting factor, which can induce a charged PC phase. Recall that chiral imbalance, i.e. a nonzero difference between densities of left- and right-handed fermions, may arise from the chiral anomaly in the quark-gluon-plasma phase of QCD and possibly leads to the chiral magnetic effect \cite{fukus} in heavy-ion collisions. It might be realized also in compact stars
or condensed matter systems \cite{andrianov} (see also the review \cite{ms}). 

\section{NJL$_{2}$ model 
}

We consider a (1+1)-dimensional NJL model in order to mimic the phase structure of real dense quark matter with two massless quark flavors ($u$ and $d$ quarks). Its Lagrangian, which is symmetrical under global color SU($N_c$) group, has the form
\begin{eqnarray}
&&  L=\bar q\Big [\gamma^\nu\mathrm{i}\partial_\nu
+\frac{\mu_B}{3}\gamma^0+\frac{\mu_I}2 \tau_3\gamma^0+\frac{\mu_{I5}}2 \tau_3\gamma^0\gamma^5\Big ]q+ \frac
{G}{N_c}\Big [(\bar qq)^2+(\bar q\mathrm{i}\gamma^5\vec\tau q)^2 \Big
],  \label{1}
\end{eqnarray}
where the quark field $q(x)\equiv q_{i\alpha}(x)$ is a flavor doublet
($i=1,2$ or $i=u,d$) and color $N_c$-plet ($\alpha=1,...,N_c$) as
well as a two-component Dirac spinor (the summation in (\ref{1})
over flavor, color, and spinor indices is implied); $\tau_k$
($k=1,2,3$) are Pauli matrices in two-dimensional flavor space. The Dirac $\gamma^\nu$-matrices ($\nu=0,1$) and $\gamma^5$ in (1) are matrices in 
two-dimensional spinor space,
\begin{equation}
\begin{split}
\gamma^0=
\begin{pmatrix}
0&1\\
1&0\\
\end{pmatrix};\qquad
\gamma^1=
\begin{pmatrix}
0&-1\\
1&0\\
\end{pmatrix};\qquad
\gamma^5=\gamma^0\gamma^1=
\begin{pmatrix}
1&0\\
0&{-1}\\
\end{pmatrix}.
\end{split}
\end{equation}
It is evident that the model (\ref{1}) is a generalization of the two-dimensional GN model 
with a single massless quark color $N_c$-plet to the case of two quark flavors
and additional baryon $\mu_B$-, isospin $\mu_I$- and axial isospin $\mu_{I5}$ chemical potentials. These parameters are introduced in order to describe in the framework of the model (1) quark matter with nonzero baryon $n_B$-, isospin $n_I$- and axial isospin $n_{I5}$ densities, respectively.
It is evident that Lagrangian (1), both at $\mu_{I5}=0$ and $\mu_{I5}\ne 0$, is invariant with respect to the abelian $U_B(1)$, $U_{I_3}(1)$ and $U_{AI_3}(1)$ groups, where 
\begin{eqnarray}
U_B(1):~q\to\exp (\mathrm{i}\alpha/3) q;~
U_{I_3}(1):~q\to\exp (\mathrm{i}\beta\tau_3/2) q;~
U_{AI_3}(1):~q\to\exp (\mathrm{i}
\omega\gamma^5\tau_3/2) q.
\label{2001}
\end{eqnarray}
So the quark bilinears $\frac 13\bar q\gamma^0q$, $\frac 12\bar q\gamma^0\tau^3 q$ and $\frac 12\bar q\gamma^0\gamma^5\tau^3 q$ are the zero components of corresponding conserved currents. Their ground state expectation values are just the baryon $n_B$-, isospin $n_I$- and chiral (axial) isospin $n_{I5}$ densities of quark matter, i.e. $n_B=\frac 13\vev{\bar q\gamma^0q}$, $n_I=\frac 12\vev{\bar q\gamma^0\tau^3 q}$
and $n_{I5}=\frac{1}{2}\vev{\bar q\gamma^0\gamma^5\tau^3 q}$. As usual, the quantities $n_B$, $n_I$ and $n_{I5}$ can be also found by differentiating the thermodynamic potential of the system with respect to the corresponding chemical potentials.
 We would like also to remark that, in addition to (\ref{2001}), Lagrangian (1) is invariant with respect to the electromagnetic $U_Q(1)$ group,
\begin{eqnarray}
U_Q(1):~q\to\exp (\mathrm{i}Q\alpha) q,
\label{2002}
\end{eqnarray}
where $Q={\rm diag}(2/3,-1/3)$.

To find the thermodynamic potential of the system, we use a semi-bosonized version of the Lagrangian (\ref{1}), which contains composite bosonic fields $\sigma (x)$ and $\pi_a (x)$ $(a=1,2,3)$ (in what follows, we use the notations $\mu\equiv\mu_B/3$, $\nu=\mu_I/2$ and $\nu_{5}=\mu_{I5}/2$):
\begin{eqnarray}
\widetilde L\ds &=&\bar q\Big [\gamma^\rho\mathrm{i}\partial_\rho
+\mu\gamma^0 + \nu\tau_3\gamma^0+\nu_{5}\tau_3\gamma^0\gamma^5-\sigma
-\mathrm{i}\gamma^5\pi_a\tau_a\Big ]q
 -\frac{N_c}{4G}\Big [\sigma\sigma+\pi_a\pi_a\Big ].
\label{2}
\end{eqnarray}
In (\ref{2}) the summation over repeated indices is implied.
From the Lagrangian (\ref{2}) one gets the Euler--Lagrange equations
for the bosonic fields
\begin{eqnarray}
\sigma(x)=-2\frac G{N_c}(\bar qq);~~~\pi_a (x)=-2\frac G{N_c}(\bar q
\mathrm{i}\gamma^5\tau_a q).
\label{200}
\end{eqnarray}
Note that the composite bosonic field $\pi_3 (x)$ can be identified
with the physical $\pi_0$ meson, whereas the physical $\pi^\pm
(x)$-meson fields are the following combinations of the composite
fields, 
$\pi^\pm (x)=(\pi_1 (x)\pm i\pi_2 (x))/\sqrt{2}$. 
Obviously, the semi-bosonized Lagrangian $\widetilde L$ is equivalent to the initial Lagrangian (\ref{1}) when using the equations (\ref{200}).
In this case, the order parameters are the ground state expectation values of the composite fields, i.e. $\vev{\sigma (x)}$ and $\vev{\pi_a (x)}$ $(a=1,2,3)$, the coordinates of the global minimum point of the thermodynamic potential $\Omega (\sigma,\pi_a)$ of the system. It is clear 
that if $\vev{\sigma(x)}\ne 0$ and/or $\vev{\pi_3(x)}\ne 0$, then the axial isospin $U_{AI_3}(1)$ symmetry of the model is spontaneously broken down, whereas at $\vev{\pi_1(x)}\ne 0$ and/or $\vev{\pi_2(x)}\ne 0$ we have a spontaneous breaking of the isospin $U_{I_3}(1)$ symmetry. Since in the last case the ground state expectation values (condensates) of both the fields $\pi^+(x)$ and $\pi^-(x)$ are not zero, this phase is usually called charged pion condensation (PC) phase.
In addition, it is easy to see from (\ref{200}) that the nonzero condensates $\vev{\pi_{1,2}(x)}$ (or $\vev{\pi^\pm(x)}$) are not invariant with respect to the electromagnetic $U_Q(1)$ transformations (\ref{2002}) of the flavor quark doublet. Hence in the charged PC phase the electromagnetic $U_Q(1)$ invariance of the model (1) is also broken  spontaneously, and superconductivity is an unavoidable property of this phase, one can call this phase pion superfluid one. 

Starting from the theory (\ref{2}), one obtains in the leading order
of the large $N_c$-expansion (i.e. in the one-fermion loop
approximation) the following path integral expression for the
effective action 
$$
\exp(\mathrm{i}{\cal S}_{\rm {eff}}(\sigma,\pi_a))=
  N'\int[d\bar q][dq]\exp\Bigl(\mathrm{i}\int\widetilde L\,d^2 x\Bigr),
$$
 Using the definition of thermodynamic
potential (TDP)
\begin{eqnarray}
\Omega (\sigma,\pi_a)~\equiv -\frac{{\cal S}_{\rm {eff}}(\sigma,\pi_a)}{N_c\int
d^2x}~\bigg |_{~\sigma,\pi_a=\rm {const}},
\end{eqnarray}
it is possible to obtain for the TDP $\Omega (\sigma,\pi_a)$ of the system:
\begin{eqnarray}
\Omega (M,\Delta)=\frac{M^2+\Delta^2}{4G}+\mathrm{i}\int\frac{d^2p}{(2\pi)^2}\ln
P_4(p_0),\,\,\,\,\,\,\,
\label{7}
 P_4(p_0)=
\eta^4-2a\eta^2-b\eta+c.
\end{eqnarray}
we use the notations:
$\eta=p_0+\mu$ and
\begin{eqnarray}
a&&=M^2+\Delta^2+p_1^2+\nu^2+\nu_{5}^2;~~b=8p_1\nu\nu_{5};\nonumber\\
c&&=a^2-4p_1^2(\nu^2+\nu_5^2)-4M^2\nu^2-4\Delta^2\nu_5^2-4\nu^2\nu_5^2.
\label{10}
\end{eqnarray}
To simplify the task, due to the fact that  the model is invariant under $U_{I_3}(1)\times U_{AI_3}(1)$ and the TDP depends effectively only on the two combinations $\sigma^2+\pi_3^2$ and $\pi_1^2+\pi_2^2$ of the bosonic fields, without loss of generality, we put $\pi_2=\pi_3=0$ in (\ref{7}), and the TDP (\ref{7}) is a function of only two variables, $M\equiv\sigma$ and $\Delta\equiv\pi_1$. 
Now let us discuss the properties of the TDP.
One can see that the TDP is invariant with respect to the so-called duality transformation 
\begin{eqnarray}
{\cal D}:~~~~M\longleftrightarrow \Delta,~~\nu\longleftrightarrow\nu_5.
 \label{16}
\end{eqnarray}
The TDP in the form of (\ref{7}) and (\ref{10}) is ultraviolet divergent and we need to renormalize it.

 The {\it unrenormalized}
TDP (\ref{7}) can be presented in the following form,
\begin{eqnarray}
\Omega (M,\Delta)&\equiv&\Omega^{un} (M,\Delta)=
\frac{M^2+\Delta^2}{4G}-
\int_{-\infty}^\infty\frac{dp_1}{4\pi}\Big (|p_{01}|+|p_{02}|+|p_{03}|+|p_{04}|\Big ). \label{8}
\end{eqnarray}
where the roots $p_{01}$, $p_{02}$, $p_{03}$ and $p_{04}$ of this polynomial $P_4(p_0)$
are the energies of quasiparticle or quasiantiparticle excitations of the system.

First of all, let us obtain a finite, i.e. renormalized, expression
for the TDP (\ref{8}) at $\mu=0$, $\nu=0$ and $\nu_5=0$, i.e. in
vacuum. 

This procedure consists of
two steps: (i) First of all we need to regularize the divergent integral, i.e. we suppose there that $|p_1|<\Lambda$. (ii)  Second, we must suppose also that the bare coupling constant $G$ depends on the cutoff parameter
$\Lambda$ in such a way that in the limit $\Lambda\to\infty$ one obtains a finite expression for the effective potential.

It can be established that the bare coupling $G\equiv G(\Lambda)$ has the following $\Lambda$ dependence:
\begin{eqnarray}
\frac 1{4G(\Lambda)}=\frac 1\pi\ln\frac{2\Lambda}{m}, \label{30}
\end{eqnarray}
where $m$ is a new free mass scale of the model, which appears instead of the dimensionless bare coupling constant $G$ (dimensional transmutation) and, evidently, does not depend on a normalization point, i.e. it is a renormalization
invariant quantity. 
We obtain
in the limit $\Lambda\to\infty$ the finite and renormalization invariant expression for the effective potential,
\begin{eqnarray}
V_0 (M,\Delta)&=&\frac{M^2+\Delta^2}{2\pi}\left [\ln\left (\frac{M^2+\Delta^2}{m^2}\right )-1\right ]. \label{31}
\end{eqnarray}

It is possible to show that 
in the general case the expression for the TDP has two terms, the first is the vacuum
effective potential and the second one is the correction due to non-zero chemical potentials. Only the first term is divergent and has to be renormalized.

Based on a numerical algorithm, it can be shown numerically that the global minimum point 
of the TDP can never be  of the form $(M_0\ne 0,\Delta_0\ne 0)$. Hence, in order to establish the phase portrait of the model, it is enough to study the projections $F_1(M)\equiv\Omega^{ren} (M,\Delta=0)$ and $F_2(\Delta)\equiv\Omega^{ren}(M=0,\Delta)$ of the TDP to the $M$ and $\Delta$ axes, correspondingly. 



\section{NJL$_{4}$ model}
Now let us describe the two flavored (3+1)-dimensional NJL model that can be used as effective model for QCD with several chemical potentials. Its Lagrangian 
has the form
\begin{eqnarray}
L=\bar q\Big [\gamma^\nu\mathrm{i}\partial_\nu
+\frac{\mu_B}{3}\gamma^0+\frac{\mu_I}2 \tau_3\gamma^0+\frac{\mu_{I5}}2\tau_{3} \gamma^0\gamma^5+\mu_{5} \gamma^0\gamma^5\Big ]q+ \frac
{G}{N_c}\Big [(\bar qq)^2+(\bar q\mathrm{i}\gamma^5\vec\tau q)^2 \Big
]  \label{14}
\end{eqnarray}
and describes dense baryonic matter with two massless $u$ and $d$ quarks, i.e. $q$ in is the flavor doublet, $q=(q_u,q_d)^T$, where $q_u$ and $q_d$ are four-component Dirac spinors as well as color $N_c$-plets (the summation in (\ref{14}) over flavor, color, and spinor indices is implied); $\tau_k$ ($k=1,2,3$) are Pauli matrices. The Lagrangian contains baryon $\mu_B$-, isospin $\mu_I$-, chiral isospin $\mu_{I5}$- chemical potentials as in the NJL$_{2}$ case. 

The quantities $n_B$, $n_I$ and $n_{I5}$ are densities of conserved charges, which correspond to the invariance of Lagrangian (1) with respect to the abelian $U_B(1)$, $U_{I_3}(1)$ and $U_{AI_3}(1)$
groups 

 As in the (1+1) dimensional case to find the TDP of the system, we use a semibosonized version of the Lagrangian (\ref{14})  and in the leading order of $1/N_{c}$ expansion it is possible to obtain the following expression for the TDP
\begin{eqnarray}
\Omega (M,\Delta)~
=\frac{M^2+\Delta^2}{4G}+\mathrm{i}\int\frac{d^4p}{(2\pi)^4}\ln
\Big(P_+(\eta)P_-(\eta)\Big),
\label{91}
\end{eqnarray}
$$
P_+(\eta)P_-(\eta)\equiv\big (\eta^4-2a_+\eta^2+b_+\eta+c_+\big )\big (\eta^4-2a_-\eta^2+b_-\eta+c_-\big )
,
$$
where $\eta=p_0+\mu$, $|\vec p|=\sqrt{p_1^2+p_2^2+p_3^2}$ and
\begin{eqnarray}
a_\pm&&=M^2+\Delta^2+(|\vec p|\pm\mu_{5})^2+\nu^2+\nu_{5}^2;~~b_\pm=\pm 8(|\vec p|\pm\mu_{5})\nu\nu_{5};\nonumber\\
c_\pm&&=a_\pm^2-4 \nu ^2
\left(M^2+(|\vec p|\pm\mu_{5})^2\right)-4 \nu_{5}^2 \left(\Delta ^2+(|\vec p|\pm\mu_{5})^2\right)-4\nu^{2} \nu_{5}^2.
\label{10}
\end{eqnarray}
It has been examined numerically and demonstrated that there is no mixed phase in the massless NJL model as in the (1+1) dimensional case. This circumstance significantly simplifies the investigation of the phase diagram of the model, since in this case it is enough to study only the projections $F_1(M)\equiv\Omega (M,\Delta=0)$ and $F_2(\Delta)\equiv\Omega(M=0,\Delta)$ of the TDP 
on the $M$ and $\Delta$ axes, correspondingly. 


Let us talk a little bit more about dualities in this case.



It is clear that the TDP of the system is invariant with respect to the transformation
\begin{eqnarray}
{\cal D}:~~~~M\longleftrightarrow \Delta,~~\nu\longleftrightarrow\nu_{5}
 \label{164}
\end{eqnarray}
 The same duality 
 was discussed in (1+1)-dimensional case. 

Let us elaborate a little more 
on the duality notion. Here it is a symmetry relation between condensates (phases) and matter content (chemical potentials). But the notion of duality is more widespread and it is a very powerful
concept that is used in different domains ranging from string theory to condensed matter physics. For example, there is a class of dualities called strong-weak dualities that connect weak coupling regime of one theory with strong coupling regime of the other. To this class belongs famous AdS/CFT (or gauge/gravity) duality \cite{Maldacena:1997re}. 
There is another class of dualities, they are called strong-strong dualities or usually bear another name large-$N_{c}$ orbifold equivalences \cite{Hanada:2011jb,
Kashiwa:2017yvy}.
Orbifold equivalences connect gauge theories with different gauge groups and matter content in the large-$N_{c}$ limit. In the framework of orbifold equivalence formalism in \cite{Hanada:2011jb} there have been also obtained a similar (to our) dualities. 

In addition to $G$ the cutoff parameter $\Lambda$ is introduced.
In the following we will use a special set of the model parameters,
$$
G=15.03\, GeV^{-1},\,\,\,\,\,\,\,\,\,\,\,\,\,\,\,\,\Lambda=0.65\, GeV.
$$
 The same parameter set has been used, e.g., in Refs \cite{Buballa:2003qv,eklim}. 
There can exist no more than three different phases in the model (1). The first one is the symmetric phase, 
where $(M_0=0,\Delta_0=0)$, 
the CSB phase, 
$(M_0\ne 0,\Delta_0=0)$ and the charged PC phase, 
$(M_0=0,\Delta_0\ne 0)$.

\section{Phase structure}

Let us consider the phase portrait of the dense baryon matter with isospin and chiral asymmetry in terms of NJL$_2$ and NJL$_4$ models.
First of all, we will study the phase structure of the NJL$_2$ model at different fixed values of the chiral isospin chemical potential $\nu_5$. Moreover, it is possible to find the
quark number density $n_q$ or baryon density $n_B$ (note that
$n_q=3n_B$). As a result, at Figs 2 left and 3 left we have
drawn several $(\nu,\mu)$-phase portraits, corresponding to
$\nu_5=0$ and $\nu_5=m$, respectively. Recall that $m$ is a free renormalization invariant mass scale
parameter, which appears in the vacuum case of the model after renormalization.

It is clear from Fig. 2 left that at
$\nu_5=0$ the charged PC phase with {\it nonzero baryon density} $n_B$
(it is denoted by the symbol PC$_{d}$) is not realized in the model under consideration. Only the charged PC phase with {\it zero baryon density} can be observed at rather small values of $\mu$. (Physically, it means that at $\nu_5=0$ the model predicts the charged PC phenomenon in the medium with $n_B=0$ only. For example, it might consist of charged pions, etc. But in quark matter with nonzero baryon density the charged PC is forbidden.) Instead, at large values of $\mu$ there exist two phases, the chiral symmetry breaking and the symmetrical one, both with nonzero baryon density, i.e. the model predicts the CSB phase of dense quark matter. However, as we can see from Fig. 3 left, at rather high values of $\nu_5$ there might appear on the phase portrait a charged PC phase with {\it nonzero baryon density}
. Hence, in chirally asymmetric, i.e. for $\nu_5>0$, and dense quark matter the charged PC phenomenon
is allowed to exist in the framework of the toy model (1). Thus, we see that $\nu_5\ne 0$ is a factor which promotes the charged PC phenomenon in dense quark matter. Note that the compact region of the $(\nu,\mu)$-plane, which is occupied by the PC$_{d}$ phase (see, e.g., Fig. 3 left at $\nu_5=m$), continues to move up along the $\mu$-axis, when $\nu_5$ increases above the value $\nu_5=m$.

Now let us consider the $(\nu,\nu_5)$-phase diagrams of the NJL$_{2}$ model. The phase diagrams  is presented at Fig. 4 left and once more support the main conclusion that the charged PC phase with nonzero baryon density, i.e. the phase denoted as PCd, might be realized in the framework of the NJL$_{2}$ model only at $\nu_5>0$. It is also clear 
that this diagram is a self-dual, i.e. the CSB and charged PC phases are arranged symmetrically with respect to the line $\nu=\nu_5$ of the $(\nu,\nu_5)$-plane.

Schematic plot of the $(\nu,\nu_5,\mu)$-parameter space phase diagram of the model is presented at Fig. 1. One can see that chiral isospin $\nu_5$ promotes the charged PC phenomenon in dense quark matter.

Now let us talk about duality ${\cal D}$ of the TDP and the phase diagram.
One can see that under the duality transformation ($\nu\leftrightarrow\nu_5$, CSB$\leftrightarrow$charged PC) $(\nu,\nu_5,\mu)$- phase
portrait is mapped to itself, i.e. the most general $(\nu,\nu_5,\mu)$-phase portrait
is self-dual.

Now let us confabulate about the phase diagram of the NJL$4$ model.
The $(\nu,\mu)$-phase diagrams of the model at different typical values of $\nu_{5}$ are presented at Figs 2 right and 3 right. One can see that  this phase diagrams look very similar to the ones corresponding to the NJL$_2$ model (Fig 2  left and 3 left). 
In the NJL$_{4}$ model case there appears a bar of CSB phase that starts from PC phase and goes along the line $\mu=\nu$. This bar-like phase can be observed at the figure of NJL$_{2}$ model case. 
At bigger values of $\nu_{5}>0$, CSB phase starts to appear from zero values of $\nu$, and the phase transition from it to the charged PC phase takes place at $\nu=\nu_{5}$ (see Fig. 2 right). As a result, the charged PC$_d$ phase is shifted to greater values of $\nu$ as one increase $\nu_{5}$. The same happens in the case of NJL$_{2}$ model, the Fig 2 left shows the $(\nu,\mu)$- phase diagram at $\nu_5=0$, if $\nu_5\neq0$ then the bar of CSB phase will shrink a little and there appears the CSB phase from zero values of $\nu$ and the phase transition from it to the charged PC phase will take place at $\nu=\nu_{5}$.
Then, at even larger values of $\nu_{5}$ the bar of CSB phase disappears (as in the  NJL$_{2}$ model case, see Fig. 3 left) and charged PC$_d$ phase shifts to the region of larger $\mu$ (see Fig. 3 right). In this region of $\nu_5$ the shape of the charged PC$_d$ phase resembles again a sole of a boot and points towards the value of $\mu=\nu_{5}$. Qualitatively same behavior can be seen in the NJL$_{2}$ model case, see Fig. 3 left.
The only qualitative difference is that in NJL$_4$ model the charged PC$_{d}$ phase is realized only for nonzero values of $\nu$ (see Fig. 3 right), whereas in NJL$_{2}$ model it takes place even for $\nu=0$ (see Fig. 3 left). But overall qualitatively behavior is the same. So one can say that in both models chiral imbalance generates charged PC phase in dense ($n_B\ne 0$) quark matter.


Now let us turn to the $(\nu,\nu_{5})$-phase in the NJL${4}$ model,
 it is depicted at Figs 4 right. 
  It can be noted that charged PC$_{d}$ and CSB$_{d}$ phases at this figure take the form of soles of boots that look towards each other and points at values of $\nu_{5}=\mu$ and $\nu=\mu$, correspondingly. Moreover, with an increase of $\mu$ charged PC$_{d}$ phase and CSB$_{d}$ phase move to the region of larger $\nu_{5}$ and $\nu$, correspondingly. 
So one can  see that nonzero $\nu_{5}$ also generates the charged pion condensation phase in dense, $n_{B}\ne 0$, quark matter in the case of NJL$_{4}$ model consideration. 
In (1+1)-dimensional case the charged PC$_d$ phase also goes to higher values of $\nu_{5}$ with increase of $\mu$ but this phase starts at zero values of $\nu$ (see Fig. 4 left), whereas in (3+1)-dimensional case the phase starts at some nonzero value of $\nu$ around $0.1$ GeV (see Fig. 4 right). So in order to realize the charged PC$_{d}$ phase in NJL$_4$ model, besides chiral imbalance there has to be the isotopic imbalance in the system. In that respect one can say that charged PC$_{d}$ phase is generated by both isospin and chiral isospin chemical potentials. This is a feature that exists only in the NJL$_4$ model, in the NJL$_{2}$ model charged pion condensate phase with nonzero baryon density could be realized just by chiral isospin chemical potential even at $\nu=0$.


\begin{figure}
\includegraphics[width=0.5\textwidth]{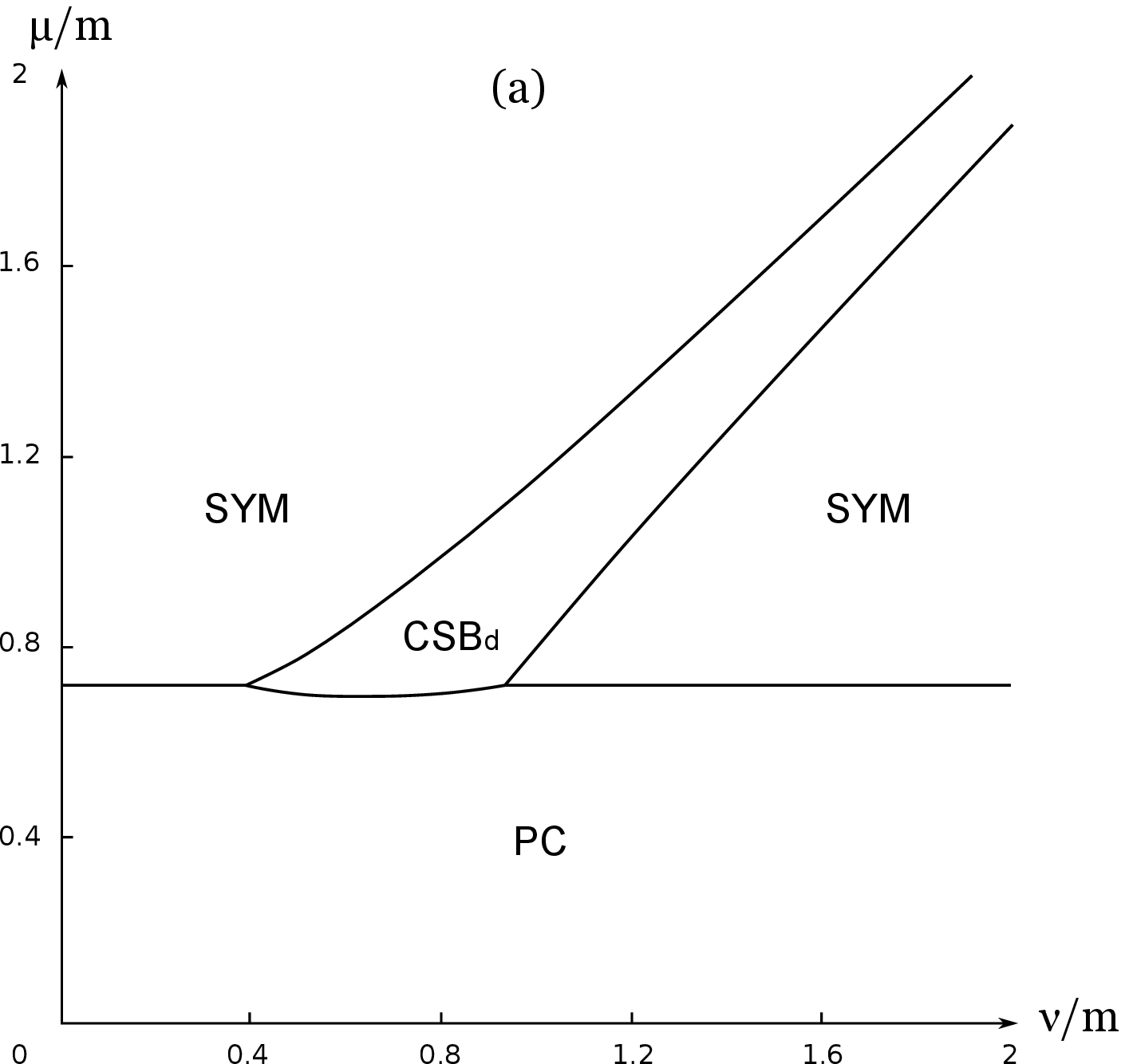}
\hfill
\includegraphics[width=0.5\textwidth]{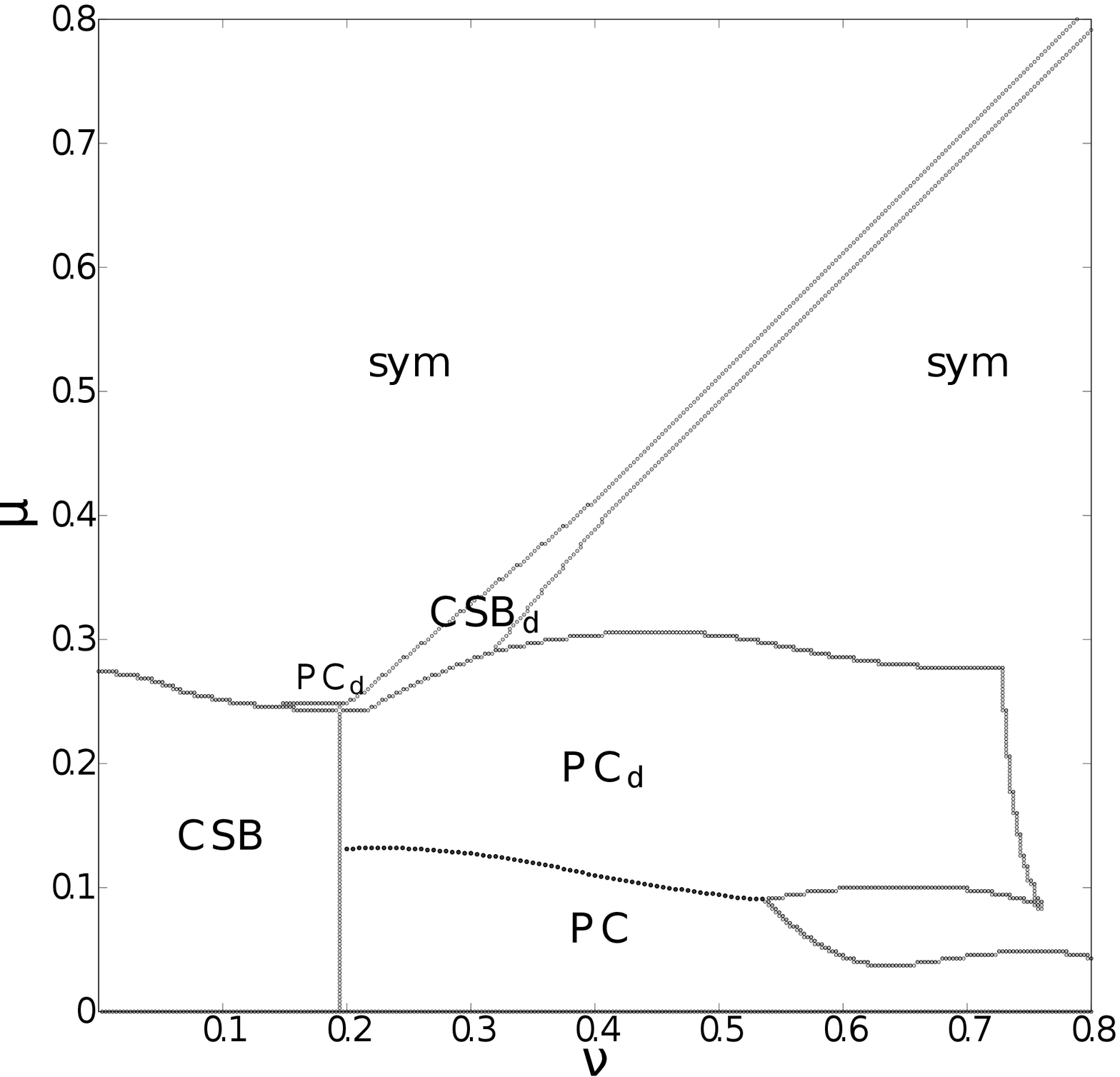}
\caption{The $(\nu,\mu)$-phase portrait of the NJL$_{2}$ and NJL$_{4}$ models: left -- the NJL$_{2}$ model case at
$\nu_5=0$. right -- the NJL$_{4}$ model case at $\nu_5=195$ MeV. 
}
\end{figure}

\begin{figure}
\includegraphics[width=0.5\textwidth]{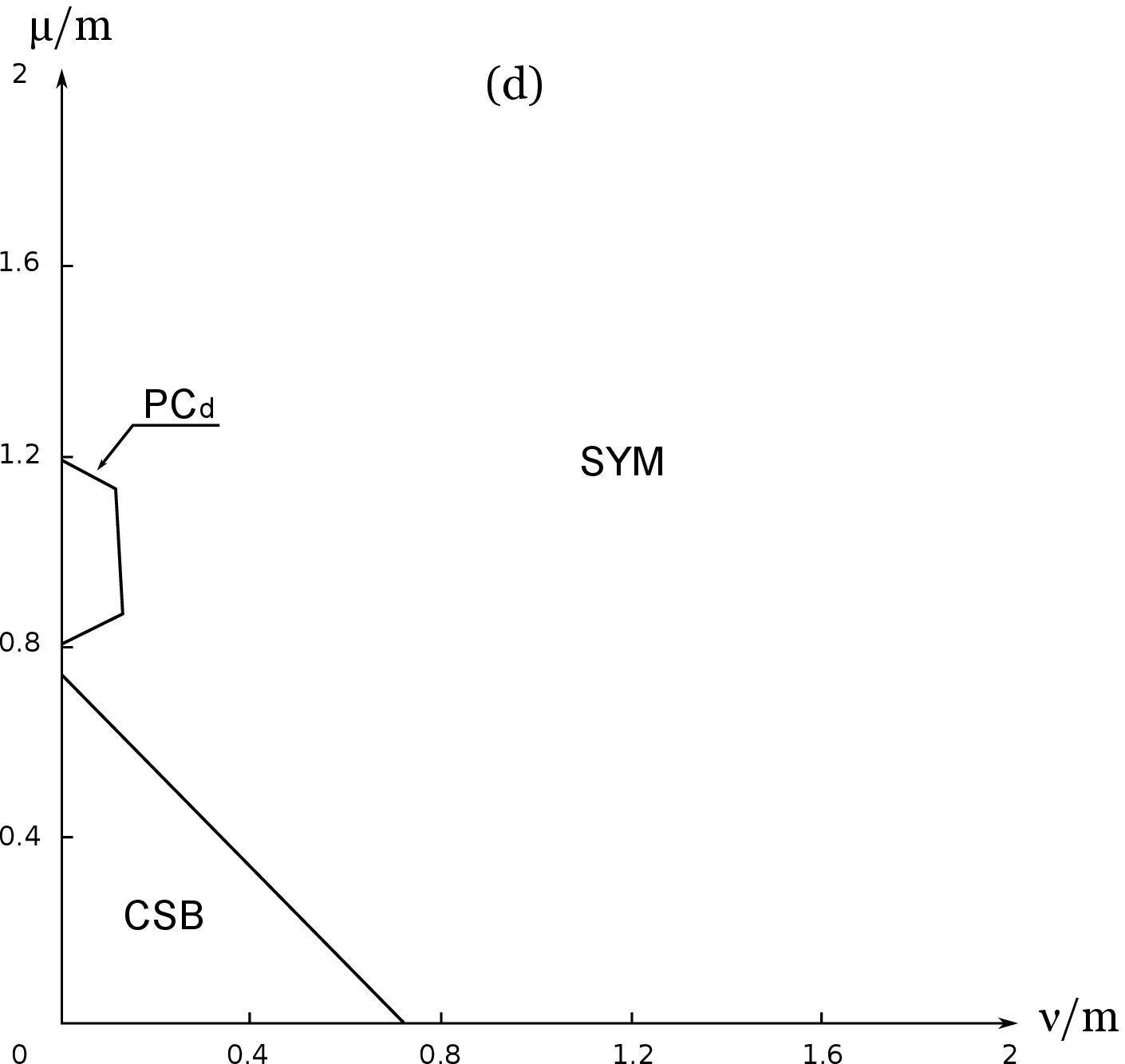}
\hfill
\includegraphics[width=0.5\textwidth]{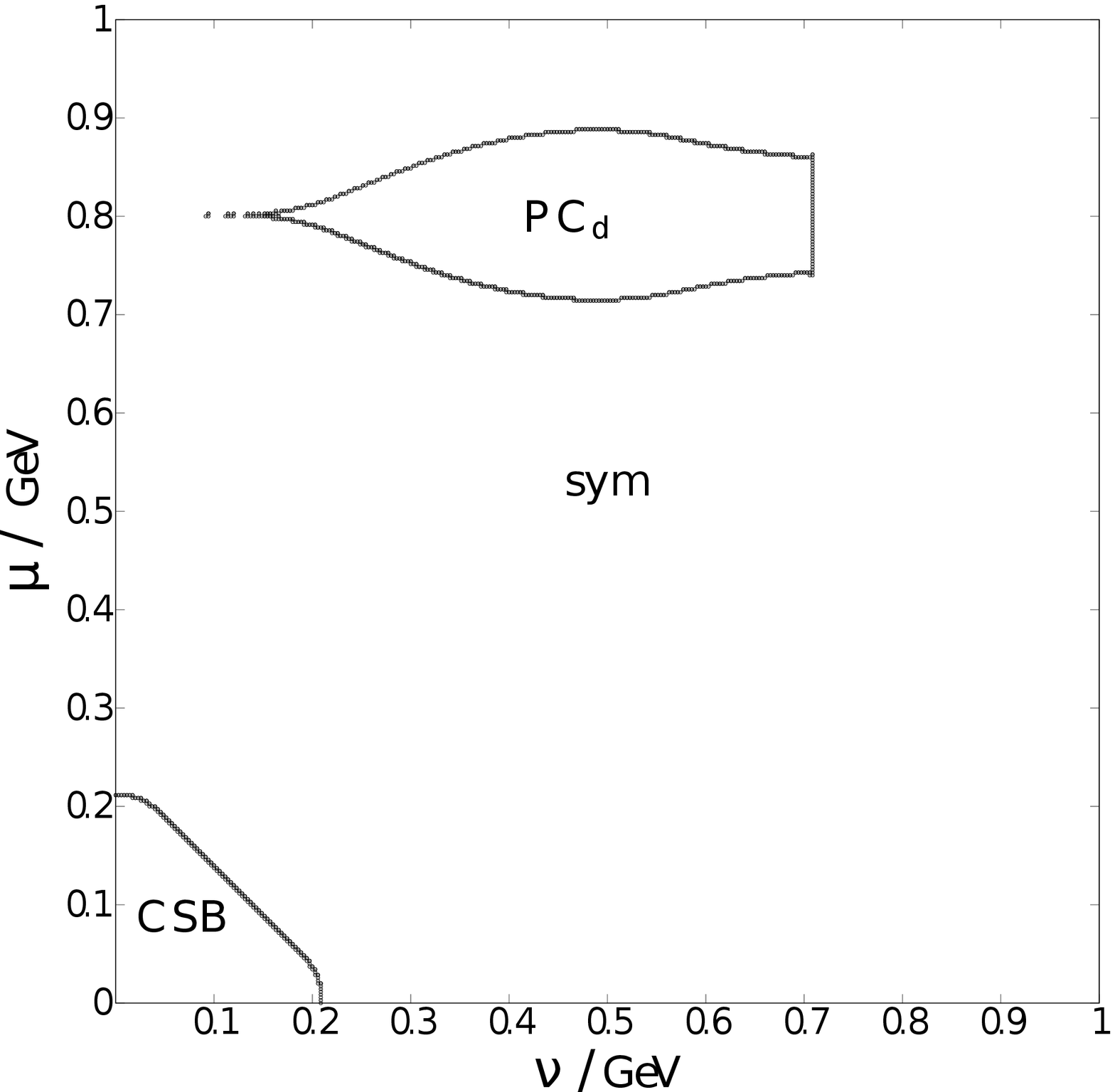}
\caption{The $(\nu,\mu)$-phase portrait of the NJL$_{2}$ and NJL$_{4}$ models: left -- the NJL$_{2}$ model case at
$\nu_5=m$. right -- the NJL$_{4}$ model case at $\nu_5=0.8$ GeV. The notations are the same.}
\end{figure}

\begin{figure}
\includegraphics[width=0.5\textwidth]{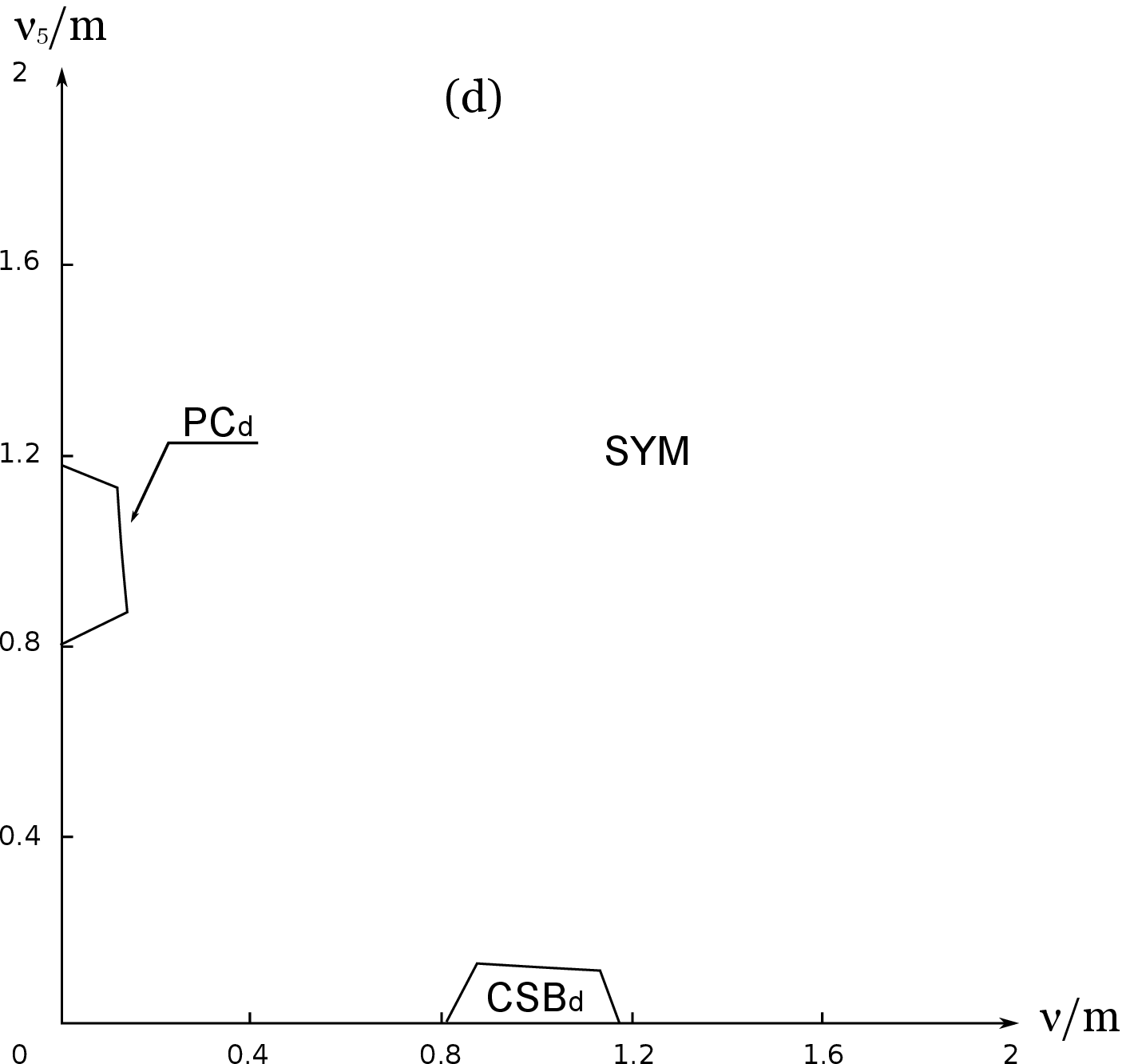}
\hfill
\includegraphics[width=0.5\textwidth]{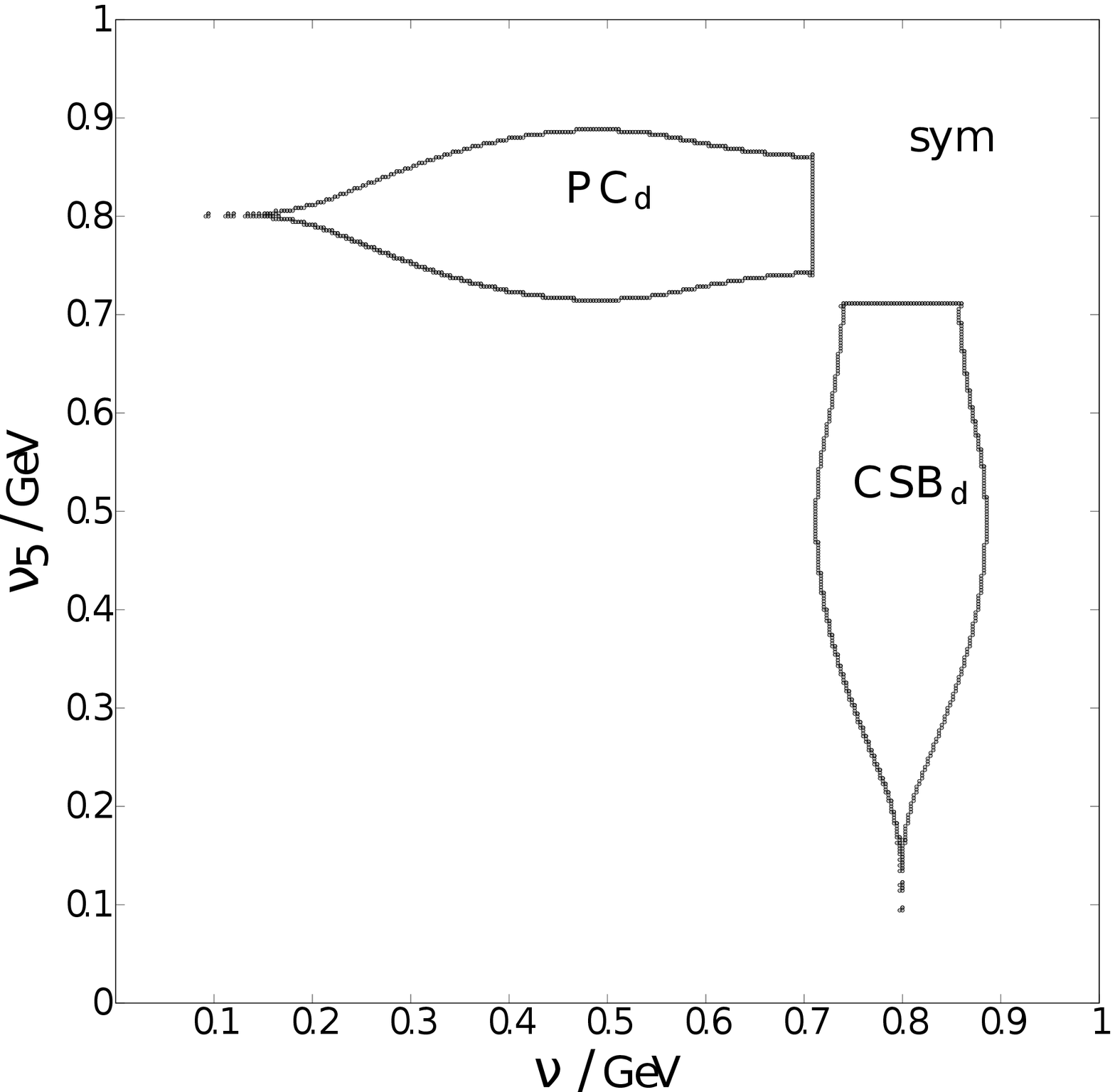}
\caption{The $(\nu,\nu_5)$-phase portrait of the NJL$_{2}$ and NJL$_{4}$ models: left -- the NJL$_{2}$ model case at
$\mu=m$. right -- the NJL$_{4}$ model case at $\mu=0.8$ GeV. The notations are the same.}
\end{figure}

\section{Summary and conclusions}

In this paper, the phase structure of the NJL$_2$ and NJL$_{4}$ models with two quark flavors is investigated in the large-$N_c$ limit in the presence of baryon
$\mu_B$, isospin $\mu_I$ and chiral isospin $\mu_{I5}$ chemical
potentials.

1) It was demonstrated in the framework of both models that chiral asymmetry (i.e. $\mu_{I5}\ne 0$)
of dense quark matter can serve as a factor promoting there a charged
pion condensation phenomenon. So this phenomenon is predicted in two models, in NJL$_2$ and NJL$_{4}$, and might be
the property of real QCD.

2) We have shown in the leading order of the large-$N_c$ approximation
that in the framework of the NJL$_2$ and NJL$_{4}$ models there is a duality
correspondence between CSB and charged PC phenomena. 

3) In contrast to  NJL$_2$ model, where the generation of the PC$_{d}$ phase occurs even at very small values of isospin chemical potential, the generation of the PC$_{d}$ phase in NJL$_{4}$  model requires not very large but nonzero isospin chemical potential.
  
  Studies in the framework of NJL$_2$ and more realistic (3+1)-dimensional NJL model in a sense complement each
other. One can consider the NJL$_2$ model at arbitrary high values of chemical potentials, whereas the NJL$_4$ model is a more realistic theory for QCD in the region of its validity. 

 Moreover, we hope that our results might shed new light
on phase structure of dense quark matter with isotopic and chiral imbalance and hence could be of importance for
describing physics in the heavy ion collision experiments and interior of neutron stars.


\begin{thebibliography}{}
%
%

\bibitem{njl}
Y. Nambu and G. Jona-Lasinio, Phys. Rev. {\bf 112}, 345 (1961).


\bibitem{Klevansky:1992qe}
S.~P.~Klevansky,
  Rev.\ Mod.\ Phys.\  {\bf 64}, 649  (1992).

\bibitem{Hatsuda:1994pi}
T.~Hatsuda and T.~Kunihiro,
  Phys.\ Rept.\  {\bf 247}, 221  (1994).

\bibitem{Buballa:2003qv}
  M.~Buballa,
Phys.\ Rept.\  {\bf 407}, 205 (2005).


\bibitem{asakawa}
M.~Asakawa and K.~Yazaki, Nucl.\ Phys.\ A {\bf 504}, 668 (1989); D.
Ebert, H. Reinhardt and M.K. Volkov, Prog. Part. Nucl. Phys. {\bf
33}, 1 (1994).

\bibitem{ebert}
D.~Ebert, K.G.~Klimenko, M.A.~Vdovichenko and A.S.~Vshivtsev,
  Phys.\ Rev.\  D {\bf 61}, 025005 (2000);
D.~Ebert and K.G.~Klimenko, Nucl.\ Phys.\  A {\bf 728}, 203 (2003).

\bibitem{sadooghi}
D.P.~Menezes, M.B.~Pinto, S.S.~Avancini, A.P.~Martinez and
  C.~Providencia, Phys.\ Rev.\  C {\bf 79}, 035807 (2009);
A.~Ayala, A.~Bashir, A.~Raya and A.~Sanchez,
  Phys.\ Rev.\  D {\bf 80}, 036005 (2009);
 S.~Fayazbakhsh and N.~Sadooghi,
  Phys.\ Rev.\ D {\bf 90}, 105030 (2014);
 E.J.~Ferrer, V.~de la Incera, J.P.~Keith, I.~Portillo and P.P.~Springsteen,
  Phys.\ Rev.\  C {\bf 82}, 065802 (2010).



\bibitem{alford}
M. Buballa, Phys. Rep. {\bf 407}, 205 (2005); I.A. Shovkovy, Found.
Phys. {\bf 35}, 1309 (2005);
 M.G.~Alford, A.~Schmitt, K.~Rajagopal, and T.~Sch\"afer,
 Rev.\ Mod.\ Phys.\  {\bf 80}, 1455 (2008).

\bibitem{klim}
D.~Ebert, V.V.~Khudyakov, V.C.~Zhukovsky and K.G.~Klimenko,
  JETP Lett.\  {\bf 74}, 523 (2001);
  Phys.\ Rev.\  D {\bf 65}, 054024 (2002);
D.~Blaschke, D.~Ebert, K.G.~Klimenko, M.K.~Volkov and V.L.~Yudichev,
  Phys.\ Rev.\  D {\bf 70}, 014006 (2004);
T.~Fujihara, D.~Kimura, T.~Inagaki and A.~Kvinikhidze,
  Phys.\ Rev.\  D {\bf 79}, 096008 (2009).

\bibitem{incera}
E.J.~Ferrer and V.~de la Incera,
  Phys.\ Rev.\  D {\bf 76}, 045011 (2007);
S.~Fayazbakhsh and N.~Sadooghi,
  Phys.\ Rev.\  D {\bf 82}, 045010 (2010);
 Phys.\ Rev.\  D {\bf 83}, 025026 (2011).


\bibitem{eklim}
D. Ebert and K.G. Klimenko, J.\ Phys.\ G {\bf 32}, 599 (2006);
Eur.\ Phys.\ J.\  C {\bf 46}, 771 (2006).

\bibitem{ak}
J.O.~Andersen and T.~Brauner,
  Phys.\ Rev.\  D {\bf 78}, 014030 (2008);
J.O.~Andersen and L.~Kyllingstad,
 J.\ Phys.\ G {\bf 37}, 015003 (2009);
 Y.~Jiang, K.~Ren, T.~Xia and P.~Zhuang,
  arXiv:1104.0094.

\bibitem{mu}
C.f.~Mu, L.y.~He and Y.x.~Liu,
  Phys.\ Rev.\  D {\bf 82}, 056006 (2010).

\bibitem{andersen}
H.~Abuki, R.~Anglani, R.~Gatto, G.~Nardulli and M.~Ruggieri,
  Phys.\ Rev.\  D {\bf 78}, 034034 (2008);
H.~Abuki, R.~Anglani, R.~Gatto, M.~Pellicoro and M.~Ruggieri,
   Phys.\ Rev.\  D {\bf 79}, 034032 (2009);
 R.~Anglani,
  Acta Phys.\ Polon.\ Supp.\  {\bf 3}, 735 (2010).

\bibitem{ekkz}
 D.~Ebert, T.G.~Khunjua, K.G.~Klimenko and V.C.~Zhukovsky,
  Int.\ J.\ Mod.\ Phys.\ A {\bf 27}, 1250162 (2012);
  Phys.\ Atom.\ Nucl.\  {\bf 77}, 795 (2014)
  [Yad.\ Fiz.\  {\bf 77}, 839 (2014)].

\bibitem{gkkz}
  N.V.~Gubina, K.G.~Klimenko, S.G.~Kurbanov and V.C.~Zhukovsky,
  Phys.\ Rev.\ D {\bf 86}, 085011 (2012).

  

\bibitem{fukus}
K. Fukushima, D.E. Kharzeev and H.J. Warringa, Phys.Rev.D {\bf 78}, 074033 (2008).

\bibitem{andrianov}
 A.A.~Andrianov, D.~Espriu and X.~Planells,
  Eur.\ Phys.\ J.\ C {\bf 73}, 2294 (2013);
 Eur.\ Phys.\ J.\ C {\bf 74}, 2776 (2014);
R.~Gatto and M.~Ruggieri,
  Phys.\ Rev.\ D {\bf 85}, 054013 (2012);
 L.~Yu, H.~Liu and M.~Huang,
 Phys.\ Rev.\ D {\bf 90}, 074009 (2014);
L.~Yu, H.~Liu and M.~Huang,
 Phys.\ Rev.\ D {\bf 94}, 014026 (2016);
G.~Cao and P.~Zhuang,
  Phys.\ Rev.\ D {\bf 92}, 105030 (2015);
V.V.~Braguta and A.Y.~Kotov,
 Phys.\ Rev.\ D {\bf 93}, no. 10, 105025 (2016);
   M.~Ruggieri and G.~X.~Peng,
  arXiv:1602.05250 [hep-ph].

\bibitem{ms}
  V.A.~Miransky and I.A.~Shovkovy,
  Phys.\ Rept.\  {\bf 576}, 1 (2015).



\bibitem{wolff}
U. Wolff, Phys. Lett. B {\bf 157}, 303 (1985);
T. Inagaki, T. Kouno, and T. Muta, Int. J. Mod. Phys. A {\bf 10}, 2241 (1995); S. Kanemura and H.-T. Sato, Mod. Phys. Lett. A {\bf 10}, 1777 (1995).

\bibitem{kgk1}
K.G. Klimenko, Theor.\ Math.\ Phys.\  {\bf 75}, 487 (1988).


\bibitem{chodos}
 A.~Chodos, H.~Minakata, F.~Cooper, A.~Singh, and W.~Mao,
  Phys. Rev. D {\bf 61}, 045011 (2000);
 K.~Ohwa, Phys.\ Rev.\  D {\bf 65}, 085040 (2002).



\bibitem{ektz}
D.~Ebert, K.G.~Klimenko, A.V.~Tyukov and V.C.~Zhukovsky,
  Phys.\ Rev.\  D {\bf 78}, 045008 (2008).

\bibitem{massive}
D.~Ebert and K.G.~Klimenko,
  Phys.\ Rev.\  D {\bf 80}, 125013 (2009);
 V.C.~Zhukovsky, K.G.~Klimenko and T.G.~Khunjua,
  Moscow Univ.\ Phys.\ Bull.\  {\bf 65}, 21 (2010).

\bibitem{ek2}
D.~Ebert and K.G.~Klimenko,
  arXiv:0902.1861 [hep-ph].

\bibitem{gubina}
D.~Ebert, N.V.~Gubina, K.G.~Klimenko, S.G.~Kurbanov and V.C.~Zhukovsky,
   Phys.\ Rev.\  D {\bf 84}, 025004 (2011).

\bibitem{Maldacena:1997re}
  J.~M.~Maldacena,
  Int.\ J.\ Theor.\ Phys.\  {\bf 38}, 1113 (1999)
[Adv.\ Theor.\ Math.\ Phys.\  {\bf 2} (1998) 231].


\bibitem{Kashiwa:2017yvy}
  K.~Kashiwa and A.~Ohnishi,
  Phys.\ Lett.\ B {\bf 772}, 669 (2017);
  M.~Hanada and N.~Yamamoto,
  JHEP {\bf 1202}, 138 (2012).

\bibitem{Hanada:2011jb} 
  M.~Hanada and N.~Yamamoto,
  PoS LATTICE {\bf 2011}, 221 (2011)
  [arXiv:1111.3391 [hep-lat]].


\end{thebibliography}
\end{document}